\documentclass[aps,prb,twocolumn,superscriptaddress]{revtex4-2}
\usepackage{amsmath,amssymb}
\usepackage{graphicx,color}
\usepackage{hyperref}
\newcommand{\nn}{\nonumber}
\newcommand{\Feps}{\mathbf{F}_{\epsilon}}
\newcommand{\dunderlines}[1]{\underline{\underline{#1}}}
\renewcommand{\vec}[1]{\underline{#1}}
\begin{document}
\title{A group theoretical approach to elasticity under constraints and predeformations}
\author{Segun Goh}
\email{s.goh@fz-juelich.de}
\affiliation{Theoretical Physics of Living Matter, Institute of Biological Information Processing, Forschungszentrum J{\"u}lich, 52425 J{\"u}lich, Germany}
\affiliation{Institut f\"ur Theoretische Physik II: Weiche Materie, Heinrich-Heine-Universit\"at D\"usseldorf, Universit{\"a}tsstra{\ss}e 1, 40225 D\"usseldorf, Germany}

\author{Hartmut L{\"o}wen}
\email{hlowen@hhu.de}
\affiliation{Institut f\"ur Theoretische Physik II: Weiche Materie, Heinrich-Heine-Universit\"at D\"usseldorf, Universit{\"a}tsstra{\ss}e 1, 40225 D\"usseldorf, Germany}

\author{Andreas M. Menzel}
\email{a.menzel@ovgu.de}
\affiliation{Institut f\"ur Physik, Otto-von-Guericke-Universit\"at Magdeburg, Universit{\"a}tsplatz 2, 39106 Magdeburg, Germany}
\date{\today}
\begin{abstract}
Respecting deformational constraints and predeformations poses a substantial challenge in the description of nonlinear elasticity. We here outline how group theory can play a beneficial role to overcome this challenge. Specifically, group theory guides us to generalized definitions of nonlinear shear deformation gradients and expressions of generalized elastic moduli in the nonlinear regime. Particularly, such achievements become important in the context of larger deformations under constraints and additional deformations on top of predeformations. \\
%highly deformable substances, for example, soft, active, or biological matter. \\ 
\end{abstract}

\maketitle

Finite deformations of solid materials require quantification in terms of nonlinear elasticity \cite{Ogden1984,Truesdell2004,Bonet2016}.
Ubiquitous examples of elastic materials that are commonly exposed to significant strains include, but are not limited to, strained-layer semiconductor heterostructures~\cite{OReilly1989,Frogley2000}, metallic alloys~\cite{Banerjee2013} including gum metals~\cite{Saito2003,Zhang2009}, two-dimensional materials~\cite{Cooper2013}, in particular, monolayer graphenes~\cite{Lee2008,Cadelano2009} (for its composites see, e.g., Ref.~\cite{Sadasivuni2014}), and carbon nanotubes~\cite{Yakobson1996,Yu2000}.
In theoretical perspective, while \emph{ab initio} calculation with the aids of density functional theory and molecular dynamics simulations are among the most frequently employed numerical approaches~\cite{Yakobson1996,Lepkowski2007,Wei2009},
the Eulerian or Lagrangian strain tensors offer a continuum mechanical framework~\cite{Birch1947,Wei2009,Tanner2019}
to address nonlinear elastic behaviors of solids.
Frequently, such systems are addressed or applied under 
maintained deformational constraints or prestrains~\cite{Jain2000,Huang2006,Ni2008,OlivaLeyva2017,Gong2021}. 
Then, the considered reference state does not correspond to the natural relaxed configuration any longer.

To characterize the stress-strain relation for small superimposed deviations in strain from the current state of the material, it is imperative to determine corresponding elastic moduli. 
In a quantitative description, one is then tempted to simply superimpose % In the infinitesimal deformation limit, a linear stress-strain relation is expected, and a 
linearized forms of strain or deformation tensors \cite{Landau1986,Sadd2009} to the already deformed state.
However, even in the limit of infinitesimal superimposed strains, such linearizations in terms of linear elasticity theory imply inconsistencies. Basic examples are included below. % for illustration. 
The reason hides in the overall finite degree of deformation that changes under the additional strain, which is not fully resolved by superimposing linearized infinitesimal strains. 
%When a maintained predeformation is assumed, however, such linearization should not be applicable because overall deformation involved is surely finite.
%From a different perspective, it is associated with the fact that one cannot recover finite deformations from linear ones.
Thus, we need to identify a formulation of the problem that consistently describes small-amplitude deformations in combination with nonlinear elasticity theory. 

This conception naturally takes us to group theory. At its core, we find the linking of infinitesimal and finite elements. More precisely, finite elements are constructed (``generated'') from infinitesimal elements (``generators'') \cite{Hamermesh1962,Rossmann2006,Hall2015}.
As we demonstrate and illustrate, % in the following, 
a consistent nonlinear elasticity theory can be formulated accordingly that naturally incorporates evaluations in constrained and predeformed states. Appropriate expressions for elastic moduli in such situations are derived. 

We start with an intuitive illustration %of the linearization issue 
in two dimensions. 
For simplicity, we confine ourselves to spatially homogeneous deformations. So-called hyperelastic materials are addressed, the stress-strain relation of which derives from a strain energy density function $W(\mathbf{F})$ \cite{Ogden2001}. Here, $\mathbf{F}$ represents the deformation gradient tensor. If $\mathbf{r}$ and $\mathbf{r}'(\mathbf{r})$ denote the positions of the material elements before and during deformation, respectively, then $\mathbf{F}=\partial\mathbf{r}'/\partial\mathbf{r}$. %When requiring rotational invariance \cite{Truesdell2004}, we can further specify $W(\mathbf{F}) = W(\mathbf{F}^T \cdot \mathbf{F})$ \cite{Ogden1984}, with $^T$ and $\cdot$ indicating the transpose and matrix multiplication, respectively. 
%Besides quantifying the pure stress-strain behavior, nonlinear elasticity allows us to characterize the material response under a maintained predeformation. 
%For illustration, 
As an example, we consider a predeformation in the form of isotropic compression or dilation of amplitude $a$, 
\begin{equation}\label{eq_F0}
\mathbf{F}_0=\left(\begin{array}{cc} a&0\\0&a \end{array}\right), 
\end{equation}
$0<a<\infty$. Maintaining this predeformed state can be regarded as a constraint. We refer to the neo-Hookean energy density \cite{Bonet2016}
\begin{equation}\label{eq_neoH}
W(%\mathbf{F}^T \cdot 
\mathbf{F})= \frac{\mu}{2} \left[{\rm Tr}\, (\mathbf{F}^T \cdot \mathbf{F}) -2 \right] -\mu \ln{J} +\frac{\lambda}{2}(\ln{J})^2, 
\end{equation}
where $J=\sqrt{\mathrm{det}(\mathbf{F}^T \cdot \mathbf{F})}$, $\mu$ and $\lambda$ are the elastic Lam\'{e} coefficients, while $^T$ and $\cdot$ indicate transpose and matrix multiplication, respectively. 

If we now wish to superimpose to this predeformation a rotation by a small rotation angle $\epsilon$, one is tempted to use the linearized form \cite{deGennes1980,Sadd2009,Menzel2007}
\begin{equation}\label{eq_Finfrot}
\mathbf{F}^{\mathrm{inf}}_{\mathrm{rot}}=
\left(\begin{array}{cc}1&-\epsilon\\\epsilon&1\end{array}\right).
\end{equation}
However, we recognize that this form is insufficient under predeformation. Particularly, it leads for $a\neq1$ to an energy difference $\Delta W(\mathbf{F}^{\mathrm{inf}}_{\mathrm{rot}})=W(\mathbf{F}^{\mathrm{inf}}_{\mathrm{rot}}\cdot\mathbf{F}_0)-W(\mathbf{F}_0)$ %due to this rotation relative to the isotropic predeformed state, we obtain for $a\neq1$
\begin{equation}
\Delta W(\mathbf{F}^{\mathrm{inf}}_{\mathrm{rot}})=%W(\mathbf{F}^{\mathrm{inf}}_{\mathrm{rot}}\cdot\mathbf{F}_0)-W(\mathbf{F}_0) =
\left[\mu(a^2-1)+\lambda(\ln a^2)\right]\epsilon^2\neq0
.%\quad\mathrm{ for }\quad a\neq1. 
\end{equation}
This contradicts the actual $\Delta W=0$, which is expected for pure rigid rotations in isotropic space. %This expression can even become negative. 
We note that the  problem is solved by using instead the actual rotation matrix to nonlinear order in $\epsilon$,
\begin{equation}\label{eq_Frot}
\mathbf{F}_{\mathrm{rot}}=
\left(\begin{array}{cc}\cos\epsilon&-\sin\epsilon\\\sin\epsilon&\cos\epsilon\end{array}\right).
\end{equation}
Then correctly $\Delta W(\mathbf{F}_{\mathrm{rot}})=W(\mathbf{F}_{\mathrm{rot}}\cdot\mathbf{F}_0)-W(\mathbf{F}_0)=0$.

Here, rectification was straightforward, because the nonlinear expression of the rotation matrix is widely known. Yet, %there arises a question: 
in general, how can we find the correct nonlinear expression for the deformation gradient tensors? Obviously, this is necessary to obtain the correct result in nonlinear elasticity theory under predeformation or other external constraints.

We find that group theory provides an answer. In two dimensions, deformation gradients are represented as $2 \times 2$ invertible matrices. Their determinants are all equal to unity, if we keep the above constraint of preserved volume while the predeformation $\mathbf{F}_0$ is maintained. 
The deformation gradient tensors of strain and rotation are elements of the special linear group $\mathsf{SL}(2, \mathbb{R})$, with regular matrix multiplication and matrix inversion as group operations. 
%whose Lie algebra $\mathfrak{sl}(2, \mathbb{R})$ is the set of all traceless $2\times 2$ matrices.
The corresponding infinitesimal generators are written as
\begin{align} \label{eq:kappa_matrices}
\boldsymbol{\kappa}_1 = \left( \begin{array}{cc} 0 & -1 \\ 1 & 0 \end{array} \right), \ 
\boldsymbol{\kappa}_2 = \left( \begin{array}{cc} 0 & 1 \\ 1 & 0 \end{array} \right), \ 
\boldsymbol{\kappa}_3 = \left( \begin{array}{cc} 1 & 0 \\ 0 & -1 \end{array} \right).
\end{align}

\begin{figure}
\includegraphics[width=0.46\textwidth]{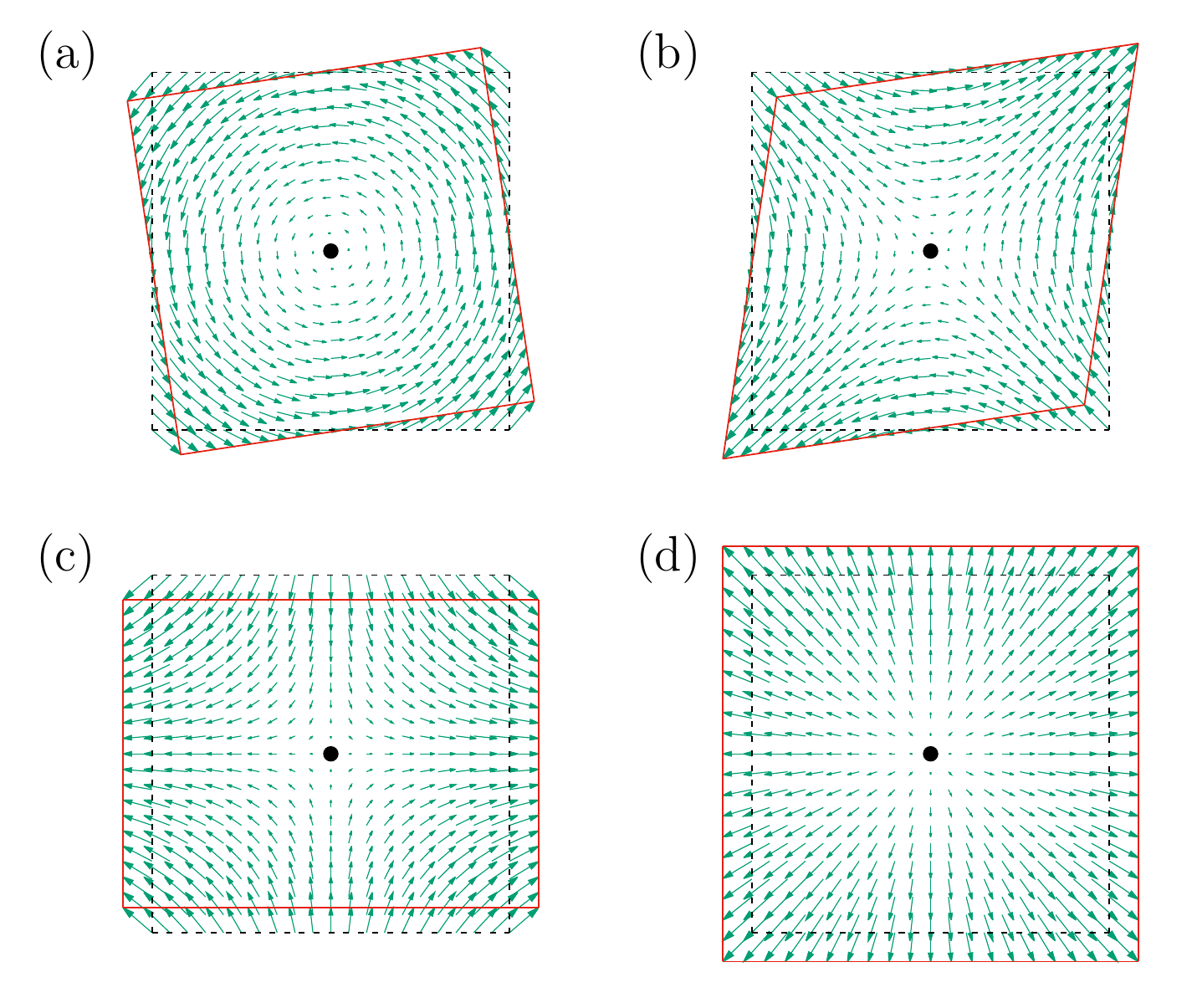}
\caption{\label{fig1} 
Geometric representation of (a) rotation, [(b) and (c)] shear deformations, and (d) dilation. 
Displacement fields (green arrows), 
undeformed (black dashed lines), and deformed (red solid lines) example systems are shown.
Black dots indicate the origin. % of each coordinate system. %The coordinates of four points at corners of the boxes representing the undeformed shape are $(-1,-1)$, $(1, -1)$, $(1,1)$, and $(-1, 1)$ (from the bottom left, in the clockwise direction).
Energetically, the shear deformations in (b) and (c) are identical for isotropic systems.}
\end{figure}

We notice that the rotation matrix $\mathbf{F}_{\mathrm{rot}}$ in Eq.~(\ref{eq_Frot}) can be obtained systematically from the generator $\boldsymbol{\kappa}_1$ as $\mathbf{F}_{\mathrm{rot}}=\exp(\epsilon\boldsymbol{\kappa}_1)$. Its action for a quadratic example system is illustrated in Fig.~\ref{fig1}(a). To linear order in $\epsilon$, we obtain $\mathbf{F}^{\mathrm{inf}}_{\mathrm{rot}}$ in Eq.~(\ref{eq_Finfrot}).

This insight guides the way to generate further deformation gradient tensors involving the other generators. For instance, $\boldsymbol{\kappa}_2$ is associated with shear. Ad hoc, we might formulate a corresponding linearized deformation gradient tensor as
\begin{equation}\label{eq_Finfshear}
\mathbf{F}^{\mathrm{inf}}_{\mathrm{shear}}=
\left(\begin{array}{cc}1&\epsilon\\\epsilon&1\end{array}\right),
\end{equation}
which indeed is useful in the absence of predeformations. However, if this deformation is superimposed to the predeformation $\mathbf{F}_0$, the energy difference $\Delta W(\mathbf{F}^{\mathrm{inf}}_{\mathrm{shear}})=W(\mathbf{F}^{\mathrm{inf}}_{\mathrm{shear}}\cdot\mathbf{F}_0)-W(\mathbf{F}_0)$ is found as
\begin{equation}
\Delta W(\mathbf{F}^{\mathrm{inf}}_{\mathrm{shear}})=%W(\mathbf{F}^{\mathrm{inf}}_{\mathrm{rot}}\cdot\mathbf{F}_0)-W(\mathbf{F}_0) =
\left[\mu(a^2+1)-\lambda(\ln a^2)\right]\epsilon^2. 
\end{equation}
This expression can even become negative for $a\neq1$, which would indicate energetically unstable situations, together with negative shear moduli. 

To find the correct nonlinear expression $\mathbf{F}_{\mathrm{shear}}$, in analogy to %Eq.~(\ref{eq_Frot}) and 
$\mathbf{F}_{\mathrm{rot}}=\exp(\epsilon\boldsymbol{\kappa}_1)$, %that enables considerations under predeformation, we in analogy to rotations 
we generate it from $\boldsymbol{\kappa}_2$ as
\begin{align} \label{eq:2D_shear}
\mathbf{F}_{\mathrm{shear}}=
\exp{(\epsilon \boldsymbol{\kappa}_2)}=\left( \begin{array}{ccc} \cosh{\epsilon} & \sinh{\epsilon} \\
\sinh{\epsilon} & \cosh{\epsilon}
\end{array} \right).
\end{align}
Indeed, we then obtain $\Delta W(\mathbf{F}_{\mathrm{shear}})=W(\mathbf{F}_{\mathrm{shear}}\cdot\mathbf{F}_0)-W(\mathbf{F}_0)=2\mu a^2\epsilon^2\geq0$. The associated shear deformation is depicted in Fig.~\ref{fig1}(b), while $\boldsymbol{\kappa}_3$ generates a shear deformation of different orientation, see Fig.~\ref{fig1}(c).
If volume changes are permitted, another generator $\boldsymbol{\kappa}_0 = \mathbf{I}$, denoting the unit matrix, needs to be added. From there, $\mathbf{F}_0$ in Eq.~(\ref{eq_F0}) can be generated, see Fig.~\ref{fig1}(d). 
Obviously, 
%As the above observation indicates, 
under constraints and finite predeformations, 
superimposed deformation gradients need to be considered to nonlinear order.
Frequently, in nonlinear theories, such constraints are handled with the aid of the method of Lagrange multipliers~\cite{Ogden1984,Holzapfel2000,Chen2003}.
Yet, this introduces additional parameters and equations. 
In Hamiltonian mechanics for particles, constraints can be eliminated from the theory  
by introducing generalized coordinates~\cite{Goldstein2001}.
However, this affords to first identify an appropriate set of generalized coordinates.
Using group theory, we here introduce a systematic way to handle nonlinear, 
finite elastic deformations, possibly subject to deformational constraints. 

Importantly, because of the constraints, % such as maintained predilation, 
the space of deformation gradient tensors $\mathbf{F}$ is not Euclidean but a manifold. 
More precisely, elasticity theory becomes based on manifolds of the general linear group $\mathsf{GL}(d,\mathbb{R})$, i.e., $\mathbf{F} \in \mathsf{GL}(d,\mathbb{R})$, where $d$ denotes the dimension.
Our approach is based on Lie algebra.

We now extend the above considerations to three dimensions. The Lie algebra $\mathfrak{gl} (3, \mathbb{R})$ of the group $\mathsf{GL}(3,\mathbb{R})$ is the set of all $3\times 3$ matrices, together with a Lie bracket operation, here the commutation relation 
$[\boldsymbol{\lambda}_i, \boldsymbol{\lambda}_j]\equiv \boldsymbol{\lambda}_i \cdot \boldsymbol{\lambda}_j -\boldsymbol{\lambda}_j \cdot \boldsymbol{\lambda}_i$.
By the set $\{\boldsymbol{\lambda}_i\}$ we denote the three-dimensional generators, here selected as~\cite{GellMann1962}
%\am{Hartmut commented on the prefactor here -- I do not have a good answer and also not on what it would mess up when changing it.} 
%\sg{SG: We need the prefactors in $\lambda_0$ and $\lambda_2$ for normalization, i.e., ${\rm Tr}\,\lambda_a \lambda_b = 2\delta_{ab}$, please see Table I, and Eqs. (4.9), (5.1), and (5.2) in Ref.~\cite{GellMann1962} by M. Gell-Mann. Maybe we should cite this paper, please see below Eq. (11).}
\begin{align} \label{eq:generator_dilation}
\boldsymbol{\lambda}_0 = \sqrt{\frac{2}{3}} \mathbf{I}, %\left( \begin{array}{ccc}
%1 & 0 & 0 \\ 0 & 1 & 0 \\ 0 & 0 & 1 \end{array} \right), \quad
\end{align}
which generates compressions or dilations, and 
\begin{align} \label{eq:generators}
&\boldsymbol{\lambda}_1 = \left( \begin{array}{ccc}
1 & 0 & 0 \\ 0 & -1 & 0 \\ 0 & 0 & 0  \end{array} \right), \quad
\boldsymbol{\lambda}_2 = \frac{1}{\sqrt{3}}\left( \begin{array}{ccc}
1 & 0 & 0 \\ 0 & 1 & 0 \\ 0 & 0 & -2  \end{array} \right), \nonumber \\ 
&\boldsymbol{\lambda}_3 = \left( \begin{array}{ccc}
0 & -1 & 0 \\ 1 & 0 & 0 \\ 0 & 0 & 0  \end{array} \right), \quad 
\boldsymbol{\lambda}_4 = \left( \begin{array}{ccc}
0 & 0 & -1 \\ 0 & 0 & 0 \\ 1 & 0 & 0  \end{array} \right), \nonumber \\
&\boldsymbol{\lambda}_5 = \left( \begin{array}{ccc}
0 & 0 & 0 \\ 0 & 0 & -1 \\ 0 & 1 & 0  \end{array} \right), \quad
\boldsymbol{\lambda}_6 = \left( \begin{array}{ccc}
0 & 1 & 0 \\ 1 & 0 & 0 \\ 0 & 0 & 0  \end{array} \right), \nn \\
&\boldsymbol{\lambda}_7 = \left( \begin{array}{ccc}
0 & 0 & 1 \\ 0 & 0 & 0 \\ 1 & 0 & 0  \end{array} \right), \quad 
\boldsymbol{\lambda}_8 = \left( \begin{array}{ccc}
0 & 0 & 0 \\ 0 & 0 & 1 \\ 0 & 1 & 0  \end{array} \right).
\end{align} 
The latter eight traceless matrices form a basis for the special linear Lie algebra $\mathfrak{sl}(3, \mathbb{R})$ \cite{Rosen1966}. %, see also Ref.~\onlinecite{Scheibner2020}.
Specifically, %in three dimensions, 
$\boldsymbol{\lambda}_1$ %is the generator of a tensile deformation 
generates stretches or compressions along the $x$ axis with compressions or stretches along the $y$ axis, respectively;
$\boldsymbol{\lambda}_2$ generates stretches or compressions along the $x$ and $y$ axes with compressions or stretches of twice the magnitude along the $z$ axis, respectively; 
$\boldsymbol{\lambda}_3$, $\boldsymbol{\lambda}_4$, and $\boldsymbol{\lambda}_5$ generate rotations in the $xy$, $xz$, and $yz$ plane, respectively;
$\boldsymbol{\lambda}_6$, $\boldsymbol{\lambda}_7$, and $\boldsymbol{\lambda}_8$ generate shear deformations in the $xy$, $xz$, and $yz$ plane, respectively.
%We note that 
$\boldsymbol{\lambda}_1$ can also be regarded to generate a shear deformation in the $xy$ plane as $\boldsymbol{\lambda}_6$, but with different orientation.

Using the exponential map~\cite{Hamermesh1962,Hall2015},
we can now generate finite deformation gradient tensors $\Feps$ from $\{\boldsymbol{\lambda}_i\}$,
\begin{align} \label{eq:small_F}
\Feps \equiv \exp{\left( \sum_{i=0}^8 \epsilon_i \boldsymbol{\lambda}_i \right)} \equiv \exp{\boldsymbol{\Lambda}},
\end{align}  
if the matrix logarithm of $\mathbf{F}_\epsilon$ exists. This is the case around $\mathbf{F}_\epsilon=\mathbf{I}$, i.e., for a set $\{\epsilon_i\}$ of finite but small coefficients.
Otherwise, we may obtain the deformation gradients from matrix multiplications of exponential maps \cite{Hall2015}, i.e.,
$\mathbf{F} = e^{\boldsymbol{\Lambda}_1}\cdot e^{\boldsymbol{\Lambda}_2} \cdot \dots\,$.
Exploiting the fact that generators provide linearly independent elements, 
a finite deformation may be decomposed into components.
For instance, based on Lie algebra, one can consider 
$\mathbf{F}_\epsilon = \exp{\left( \epsilon_0 \boldsymbol{\lambda}_0 +\epsilon_3 \boldsymbol{\lambda}_3 + \epsilon_6 \boldsymbol{\lambda}_6 \right)}$
as a superposition of
a dilation (or compression) with the strength of $\epsilon_0$, and
a rotation and a shear deformation in the $xy$ plane with strengths $\epsilon_3$ and $\epsilon_6$, 
respectively, which is distinguished from conventional decompositions  
in terms of deformation gradient tensors.
Indeed, direct decompositions of deformation gradient tensors 
in the form of $\Feps = \mathbf{F}_1 \cdot \mathbf{F}_2$, 
see, e.g., Ref.~\cite{Lee1969}, are not allowed in general, 
as generators mostly do not commute.
Rather than that, a deformation should be divided into many pieces of 
infinitesimal deformations due to the Lie product formula~\cite{Hall2015}. 

Using the generators, our next step is to derive appropriate expressions for the elastic moduli and rotation coefficients for a system in a constrained or predeformed state. 
We introduce a vector notation $\vec{\epsilon} \equiv (\epsilon_0, \ldots, \epsilon_8)^T$ and under a finite predeformation $\mathbf{F}_0$ expand $W$ in terms of $\{\epsilon_i\}$ as
\begin{align} \label{eq:expansion}
W(\Feps\cdot\mathbf{F}_0) = W_0 + \vec{s}^{\,T} \cdot \vec{\epsilon} +\frac{1}{2}\vec{\epsilon}^{\,T} \cdot \dunderlines{C} \cdot \vec{\epsilon},
\end{align}
where $W_0=W(\mathbf{F}_0)$ and for $i,j\in\{0,\ldots,8\}$ we have
\begin{align}\label{eq_sC}
s_i = \frac{\partial W}{\partial \epsilon_i} \quad \mathrm{and} \quad {C}_{ij} =\frac{\partial^2 W}{\partial \epsilon_i \partial \epsilon_j}.  %, \quad {\rm for}\quad i,j\in\{0,\ldots,8\}.
\end{align}
The vector $\vec{s}$, conjugate to $\vec{\epsilon}$,
quantifies the stress under a constraint or predeformation. 

Generally, a carefully selected basis may support the description of the problem. For instance, fixing the volume in the predeformed state can simply be achieved by omitting the component $\epsilon_0$. 
Linear algebra allows to adjust the basis to the problem at hand. 
Specifically, unitary operators $\dunderlines{U}$ 
that connect two different bases via $\vec{\tilde{\epsilon}} = \dunderlines{U} \cdot \vec{\epsilon}$ imply
\begin{equation} \label{eq:linear_transformation}
W = W_0 + \vec{\tilde{s}}^{\,T} \cdot \vec{\tilde{\epsilon}} + \frac{1}{2} \vec{\tilde{\epsilon}}^{\,T} \cdot \dunderlines{\tilde{C}} \cdot \vec{\tilde{\epsilon}},
\end{equation}
where $\vec{\tilde{s}} = \dunderlines{U} \cdot \vec{s}$ and $\dunderlines{\tilde{C}} = \dunderlines{U} \cdot \dunderlines{C} \cdot \dunderlines{U}^T$.

In what follows, we investigate the roles that $\vec{s}$ and $\dunderlines{C}$ play in the context of appropriate elastic moduli %can be expressed in terms of Cauchy stress tensors and instantaneous moduli in 
for nonlinear elasticity theory.
%The first order derivatives yield 
We use Einstein's summation convention and denote as $\boldsymbol{\sigma}$ the Cauchy stress tensor, which is given by 
\begin{align}
\boldsymbol{\sigma} = \frac{1}{J}\frac{\partial W}{\partial \mathbf{F}}\cdot \mathbf{F}^T,
\end{align}
for hyperelastic materials. 
From Eq.~(\ref{eq_sC}), we find in coordinates associated with the deformed state~\cite{Ogden1984}
\begin{align} \label{eq:cauchy_stress}
s_i %\equiv \frac{\partial W}{\partial \epsilon_i}
=  \frac{\partial W}{\partial [\Feps]_{ab}}
	\left[\frac{\partial \Feps}{\partial \epsilon_i}\right]_{ab} 
= J [\boldsymbol{\sigma}]_{ab} \left[ \frac{\partial \Feps}{\partial \epsilon_i} \cdot \Feps^{-1} \right]_{ba} . 
\end{align}
Since we are working with matrix Lie groups, we may insert the expression \cite{Rossmann2006,Hall2015}
%Confining ourselves to matrix Lie groups, the derivative term is given as~\cite{Rossmann2006,Hall2015}
\begin{align}
\frac{\partial \Feps}{\partial \epsilon_i} 
= \Feps \cdot \int_0^1 {\rm d}s\, {\rm Ad}_{e^{-s\boldsymbol{\Lambda}}} (\boldsymbol{\lambda}_i),
\end{align} 
where $\boldsymbol{\Lambda} = \sum_i \epsilon_i \boldsymbol{\lambda}_i$, ${\rm Ad}_{e^\mathbf{X}} (\mathbf{Y})= e^{\rm{ad}_\mathbf{X}} (\mathbf{Y})$, and ${\rm ad}_\mathbf{X} (\mathbf{Y}) = [\mathbf{X}, \mathbf{Y}]$.
Particularly, Eq.~(\ref{eq:cauchy_stress}) connects the newly defined first-order coefficients $\{s_i\}$ and the Cauchy stress tensor.
%In terms of group theory, the Cauchy stress tensor is associated with coefficients of the adjoint representation.
For example, we obtain
\begin{align} \label{eq:symmetric_s0_pressure}
s_0 = \sqrt{\frac{2}{3}}J\, {\rm Tr}\,\boldsymbol{\sigma} \equiv - \sqrt{6} J p
\end{align} 
because $[\boldsymbol{\lambda}_0, \boldsymbol{\Lambda}] = 0$ and consequently $\partial \Feps /\partial \epsilon_0 = \Feps \cdot \boldsymbol{\lambda}_0$, while $p$ denotes the generalized pressure.

%In the same manner, 
Analogously, we obtain for the second-order coefficients, which we now call generalized elastic moduli,  
\begin{align} \label{eq:second_order_coefficient}
{C}_{ij} %= \frac{\partial^2 W}{\partial \epsilon_i \partial \epsilon_j}
= [\boldsymbol{\mathcal{A}}]_{abcd}\! \left[\frac{\partial \Feps}{\partial \epsilon_i}\right]_{\!cd}  \!
	\left[\frac{\partial \Feps}{\partial \epsilon_j}\right]_{\!ab} 
	\!\!+J [\boldsymbol{\sigma}]_{ab}\! \left[ \frac{\partial^2 \Feps}{\partial \epsilon_i \partial \epsilon_j} \!\cdot \Feps^{-1} \right]_{\!ba}\!\!. 
\end{align}
Here, 
$\boldsymbol{\mathcal{A}}$ represents the fourth-rank tensor of classic elastic moduli with components
\begin{align} \label{def:instantaneous}
[\boldsymbol{\mathcal{A}}]_{abcd} \equiv \left. \frac{\partial^2 W}{\partial [\Feps]_{cd}\, \partial [\Feps]_{ab}}  \right|_{\epsilon \to 0}.
\end{align}
They completely determine $\dunderlines{C}$ in the absence of any predeformation, i.e., for $\boldsymbol{\sigma} = \boldsymbol{0}$.  
In this case, 
the form of $\boldsymbol{\mathcal{A}}$, and subsequently of $\dunderlines{C}$ 
is fully determined by irreducible representations in linear elasticity, 
see, e.g., Ref.~\cite{Kleinert1989}. 
Otherwise, the second-order derivatives are calculated via \cite{Rossmann2006,Hall2015}
\begin{align}\label{eq_2nd}
\frac{\partial^2 \Feps}{\partial \epsilon_i \partial \epsilon_j} 
= \Feps& \cdot \int_0^1 {\rm d}s \left[ \int_s^1{\rm d}t\, ({\rm Ad}_{e^{-t\boldsymbol{\Lambda}}}(\boldsymbol{\lambda}_j))({\rm Ad}_{e^{-s\boldsymbol{\Lambda}}}(\boldsymbol{\lambda}_i)) \right. \nonumber \\
&\left.
+\int_0^s {\rm d}t\, ({\rm Ad}_{e^{-s\boldsymbol{\Lambda}}}(\boldsymbol{\lambda}_i))({\rm Ad}_{e^{-t\boldsymbol{\Lambda}}}(\boldsymbol{\lambda}_j)) \right]. 
\end{align}
We note that our expression for $C_{ij}$ in Eq.~\eqref{eq:second_order_coefficient} corresponds to a type of tangent moduli~\cite{Bigoni2012} quantifying elastic moduli in predeformed states, 
but automatically satisfies imposed constraints 
if an appropriate set of generators is chosen.
Remarkably, the newly derived corrections 
due to imposed predeformations given by the second term in Eq.~\eqref{eq:second_order_coefficient}, 
together with $\boldsymbol{\mathcal{A}}$ associated with the ground-state symmetry, 
provide irreducible representations of nonlinear elastic moduli extended to 
general states of systems. 

For simplicity, we henceforth confine ourselves to infinitesimal volume-preserving deformations with $J = 1$.
Then, from Eqs.~\eqref{eq:small_F} and \eqref{eq:cauchy_stress} we find
\begin{align}\label{eq_si}
s_i = [\boldsymbol{\sigma}]_{ab} [ \boldsymbol{\lambda}_i]_{ba}.
\end{align}
For unconstrained systems, the Cauchy stress tensor $\boldsymbol{\sigma}_{\rm un}$ can be decomposed into 
the one for incompressible systems $\boldsymbol{\sigma}$ and an $s_0$-term as
%\begin{align}
$\boldsymbol{\sigma}_{\rm un} = \boldsymbol{\sigma} +%\frac{1}{3}
s_0 \mathbf{I}/\sqrt{6} = \boldsymbol{\sigma}-p\mathbf{I}$, see Eq.~\eqref{eq:symmetric_s0_pressure},  
%\end{align}
in line with the method of Lagrange multipliers \cite{Ogden1984}. 
%This relation simply recovers the stress tensor for incompressible systems derived from the Lagrange multiplier method (see, e.g., Ref.~\onlinecite{Ogden1984}),
%as the second term can be replaced by $-p\mathbf{I}$, see Eq.~\eqref{eq:symmetric_s0_pressure}.
Moreover, the generalized elastic moduli in Eq.~\eqref{eq:second_order_coefficient} via Eq.~\eqref{eq_2nd} reduce to  
\begin{align} \label{eq:second_order_infinitesimal}
{C}_{ij} = %\frac{\partial^2 W}{\partial \epsilon_i \partial \epsilon_j} = 
[\boldsymbol{\mathcal{A}}]_{abcd} [\boldsymbol{\lambda}_i]_{cd}[\boldsymbol{\lambda}_j]_{ab} 
	+ \frac{1}{2} [\boldsymbol{\sigma}]_{ab} [\{\boldsymbol{\lambda}_i, \boldsymbol{\lambda}_j\}]_{ba}.
\end{align}
In this expression, $\{\boldsymbol{\lambda}_i, \boldsymbol{\lambda}_j\}=\boldsymbol{\lambda}_i \cdot \boldsymbol{\lambda}_j + \boldsymbol{\lambda}_j \cdot \boldsymbol{\lambda}_i$ denote the anticommutation relations.
It is straightforward to calculate them from Eqs.~\eqref{eq:generator_dilation} and \eqref{eq:generators}. They can be rewritten in the form 
%\begin{align} \label{eq:anti1}
$\{\boldsymbol{\lambda}_i, \boldsymbol{\lambda}_j \} = g^{ij} \mathbf{I} + \sum_{k=1}^8 h^{ijk} \boldsymbol{\lambda}_k$,
%\end{align}
%with $\mathbf{I}$ denoting the identity matrix, where the coefficients
where
$g^{ij}$ and $h^{ijk}$ represent the so-called structure constants for the associated Lie algebra, here $\mathfrak{gl}(3,\mathbb{R})$, that can be calculated explicitly.
%Numerical values are listed in \textcolor{red}{Appendix}.
%follow from explicit calculation and are listed
%in Ref.~\onlinecite{SI}.}
%\textcolor{red}{Specifically,
%we find from explicit calculation
%\begin{align} \label{eq:anti2}
%g^{00}&=g^{11}=g^{22}=-g^{33}=-g^{44}=-g^{55} =g^{66}=g^{77}=g^{88} \nn \\
%	&=\frac{4}{3}
%\end{align}
%and
%\begin{align}
%&h^{011}=h^{022}=h^{033}=h^{044}=h^{055}=h^{066}=h^{077}=h^{088} \nn \\
%& = \frac{2\sqrt{6}}{3}, \\ 
%&h^{112}=h^{121}=-h^{222}=h^{233}=h^{266}=-h^{332}=h^{662}\nn \\
%&= \frac{2}{\sqrt{3}}, \\
%&h^{244}=h^{255}=h^{277}=h^{288}=-h^{442}=-h^{552}=h^{772}=h^{882} \nn \\
%&=-\frac{1}{\sqrt{3}}, \\
%&h^{144}=-h^{155}=h^{177}=-h^{188}
%= -h^{348}=h^{357}=-h^{375} \nn \\
%  &=h^{384}
%=-h^{441}=-h^{456}=h^{465}=h^{483}
%=h^{551}=h^{564} \nn \\
% &=-h^{573}=h^{678}=h^{687}
% =h^{771}=h^{786}=-h^{881}
%=1,  \label{eq:anti3} 
%\end{align}
%together with $h^{ijk} = h^{jik}$ because $\{\boldsymbol{\lambda}_i, \boldsymbol{\lambda}_j \}=\{\boldsymbol{\lambda}_j, \boldsymbol{\lambda}_i \}$. %\rev{All nonvanishing coefficients should be listed already above, i.e. also those including 0 as an index.}
%Nonlisted coefficients are zero.}
Thus we obtain from Eqs.~\eqref{eq:symmetric_s0_pressure}, \eqref{eq_si}, and \eqref{eq:second_order_infinitesimal} 
\begin{align} \label{eq:sigma_structure_constant}
{C}_{ij} = %\frac{\partial^2 W}{\partial \epsilon_i \partial \epsilon_j} = 
[\boldsymbol{\mathcal{A}}]_{abcd} [\boldsymbol{\lambda}_i]_{cd}[\boldsymbol{\lambda}_j]_{ab}
+ \frac{1}{2}\left( \sqrt{\frac{3}{2}}g^{ij} s_0 +\sum_{k=1}^8 h^{ijk} s_k \right).
\end{align}
The first contribution, in terms of the matrix components for $i,j = 1, \ldots, 8$, 
related to both prestressed and nonprestressed systems, takes the form
\begin{align}
\sqrt{\frac{2}{3}}\left( \begin{array}{cccccccc} 
s_0 & 0 & 0 & 0 & 0 & 0 & 0 & 0 \\
0 & s_0 & 0 & 0 & 0 & 0 & 0 & 0 \\
0 & 0 & -s_0 & 0 & 0 & 0 & 0 & 0 \\
0 & 0 & 0 & -s_0 & 0 & 0 & 0 & 0 \\
0 & 0 & 0 & 0 & -s_0 & 0 & 0 & 0 \\
0 & 0 & 0 & 0 & 0 & s_0 & 0 & 0 \\
0 & 0 & 0 & 0 & 0 & 0 & s_0 & 0 \\
0 & 0 & 0 & 0 & 0 & 0 & 0 & s_0
		\end{array} \right). %_{\!\!ij}
\end{align}
The square root appears because of the definition of $\boldsymbol{\lambda}_0$ 
in Eq.~(\ref{eq:generator_dilation}) that indicates the square root as a prefactor. 
The second contribution in Eq.~\eqref{eq:sigma_structure_constant}, 
which needs to be taken into account in the case of prestressed systems, reads
\begin{widetext}
\begin{eqnarray} \label{eq:second_order_infinitesimal_appendix}
\left(
\renewcommand{\arraystretch}{1.4}
\begin{array}{cccccccc} 
\frac{1}{\sqrt{3}}s_2 	& \frac{1}{\sqrt{3}}s_1 	& 0 & \frac{1}{2}s_4 & -\frac{1}{2}s_5 & 0 & \frac{1}{2}s_7 & -\frac{1}{2}s_8 \\
\frac{1}{\sqrt{3}}s_1 	& -\frac{1}{\sqrt{3}}s_2 	& \frac{1}{\sqrt{3}}s_3 & -\frac{1}{2\sqrt{3}}s_4 & -\frac{1}{2\sqrt{3}}s_5 & \frac{1}{\sqrt{3}}s_6 & -\frac{1}{2\sqrt{3}}s_7 	& -\frac{1}{2\sqrt{3}}s_8 \\
0 								& \frac{1}{\sqrt{3}}s_3 	& -\frac{1}{\sqrt{3}}s_2 & -\frac{1}{2}s_8 & \frac{1}{2}s_7	& 0 & -\frac{1}{2}s_5 & \frac{1}{2}s_4 \\
\frac{1}{2}s_4 			& -\frac{1}{2\sqrt{3}}s_4	& -\frac{1}{2}s_8 & -\frac{1}{2}s_1 + \frac{1}{2\sqrt{3}}s_2 & -\frac{1}{2}s_6 & \frac{1}{2}s_5 & 0 & \frac{1}{2}s_3 \\
-\frac{1}{2}s_5			& -\frac{1}{2\sqrt{3}}s_5	& \frac{1}{2}s_7 & -\frac{1}{2}s_6 & \frac{1}{2}s_1 +\frac{1}{2\sqrt{3}}s_2 & \frac{1}{2}s_4 & -\frac{1}{2}s_3 & 0 \\
0 								& \frac{1}{\sqrt{3}}s_6		& 0 & \frac{1}{2}s_5 & \frac{1}{2}s_4 & \frac{1}{\sqrt{3}}s_2 & \frac{1}{2}s_8 & \frac{1}{2}s_7 \\
\frac{1}{2}s_7 			& -\frac{1}{2\sqrt{3}}s_7	& -\frac{1}{2}s_5 & 0 & -\frac{1}{2}s_3 & \frac{1}{2}s_8 & \frac{1}{2}s_1 -\frac{1}{2\sqrt{3}}s_2	& \frac{1}{2}s_6 \\
-\frac{1}{2}s_8 			& -\frac{1}{2\sqrt{3}}s_8	& \frac{1}{2}s_4 & \frac{1}{2}s_3 & 0 & \frac{1}{2}s_7 & \frac{1}{2}s_6 & -\frac{1}{2}s_1 - \frac{1}{2\sqrt{3}}s_2
		\end{array} \right). %_{\!\!ij}\!\!.
%		\nn\\&&
\end{eqnarray}
\end{widetext}
Together, Eqs.~\eqref{eq:sigma_structure_constant}--\eqref{eq:second_order_infinitesimal_appendix} conclude our derivation of the generalized elastic moduli $C_{ij}$. They follow in a systematic way using group theory. Beyond the pure classic elastic moduli associated with $\boldsymbol{\mathcal{A}}$, see Eq.~\eqref{def:instantaneous}, Eq.~\eqref{eq:sigma_structure_constant} contains the contributions through the predeformation via the Cauchy stresses $\boldsymbol{\sigma}$, see Eqs.~\eqref{eq:symmetric_s0_pressure} and \eqref{eq_si}.
The tensor $\boldsymbol{\mathcal{A}}$ still refers to all modes of deformation, including the ones that actually are restricted by the imposed constraints. 
Our formalism consistently includes the consequences of these constraints into the overall expression for the generalized elastic moduli $\underline{\underline{C}}$.
Moreover, as a strong benefit, the factors $\vec{s}$ can be directly read off from an expansion of the deformation energy in a suitable basis $\vec{\epsilon}$ adjusted to the constraints, see Eqs.~\eqref{eq:expansion} and \eqref{eq:linear_transformation}. 
For situations of completely constrained volume, the contributions by $s_0$ and $\boldsymbol{\lambda}_0$ may simply be dropped. 
%

%\section{Example: neo-Hookean model revisited}
For a brief illustration of our formalism, we return to Eq.~\eqref{eq_neoH} in two dimensions and supplement it as~\cite{Toner1981, Nelson1981}
\begin{eqnarray}\label{eq:neoHookean_W}
W(\mathbf{F})&= &\frac{\mu}{2} \left[ {\rm Tr}\, (\mathbf{F}^T \cdot \mathbf{F}) -2\right] -\mu \ln{J} +\frac{\lambda}{2}(\ln{J})^2 \nonumber \\
		&&{}-\frac{1}{2} \left(H \hat{n}_H \cdot \frac{\mathbf{F}\cdot\hat{n}_s}{|\mathbf{F}\cdot\hat{n}_s|} \right)^2 .
\end{eqnarray}
The last term includes the orientation of an internal axis $\hat{n}_s$ that is reoriented by $\mathbf{F}$ and coupled to an external field $H\hat{n}_H$. % that identifies anisotropy of the system. 
%We set $\hat{n}_H = (0,1)^T$ and $\hat{n}_s = (0,1)^T$.
Together with the predeformation in Eq.~\eqref{eq_F0}, % Again, we introduce a predeformation of $\mathbf{F}_0$, i.e., 
we consider $\mathbf{F} = \Feps \cdot \mathbf{F}_0 = a \Feps$, while imposing for $\Feps$ a constraint of preserved volume. 

\begin{figure}
\includegraphics[width=0.46\textwidth]{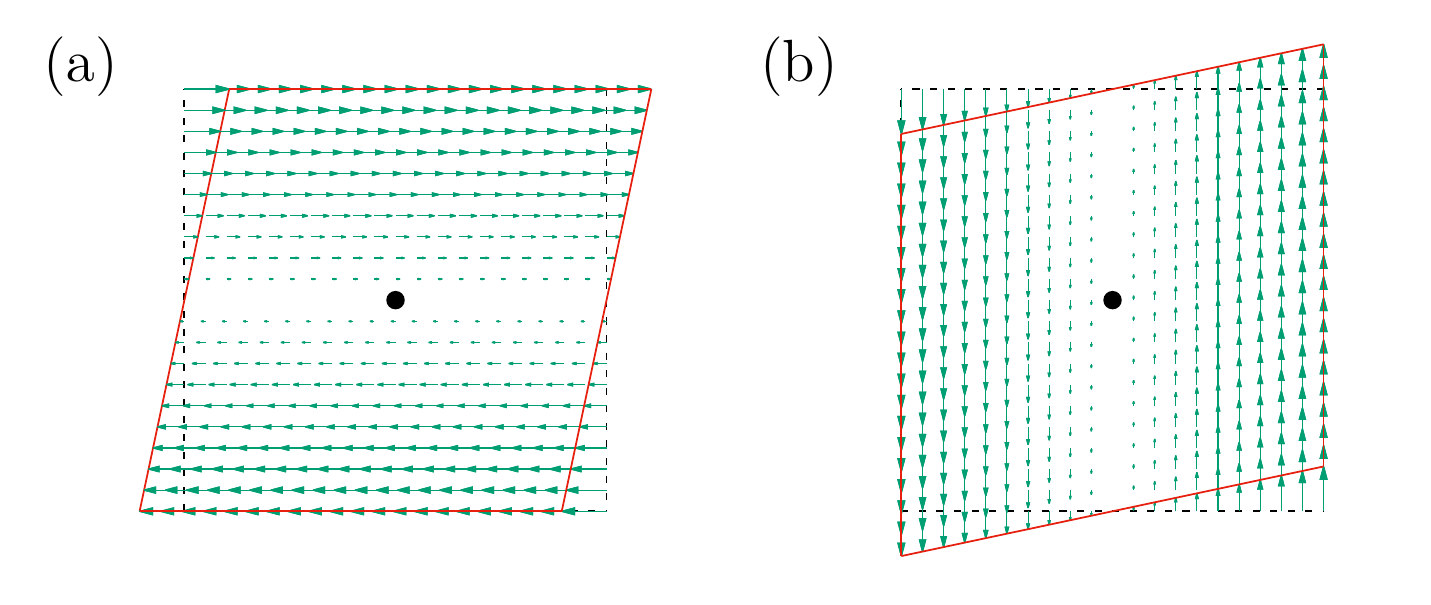}
\caption{\label{fig2} 
Geometric representation of simple shears in analogy to the illustration in Fig.~\ref{fig1}.}
%Displacement fields (green arrows), 
%undeformed (black dashed lines), and deformed (red solid lines) example systems are shown.
%Black dots indicate the origin. % of each coordinate system. %The coordinates of four points at corners of the boxes representing the undeformed shape are $(-1,-1)$, $(1, -1)$, $(1,1)$, and $(-1, 1)$ (from the bottom left, in the clockwise direction).
%Energetically, the shear deformations in (b) and (c) are identical for isotropic systems.}
\end{figure}

A second set of generators
\begin{align} \label{eq:kappa_tilde}
\tilde{\boldsymbol{\kappa}}_1 = \left( \begin{array}{cc} 0 & \sqrt{2} \\ 0 & 0 \end{array} \right), \quad
\tilde{\boldsymbol{\kappa}}_2 = \left( \begin{array}{cc} 0 & 0 \\ \sqrt{2} & 0 \end{array} \right), \quad
\tilde{\boldsymbol{\kappa}}_3 = \boldsymbol{\kappa}_3
\end{align}
is used besides Eq.~\eqref{eq:kappa_matrices}, 
where $\tilde{\boldsymbol{\kappa}}_1$ and $\tilde{\boldsymbol{\kappa}}_2$ generate frequently considered simple shears~\cite{Sadd2009}, see Fig.~\ref{fig2}(a) and (b), respectively.
%Here, the factor of $\sqrt{2}$ has been included to render the transformation between the two sets of generators unitary, see Eq.~\eqref{eq:linear_transformation}. Specifically, 
The unitary matrix %\am{If we agree on the double-headed arrows for $C$, then all $U$'s should be written in the same way, I think.}
%$\mathbf{U}$ for the transformation $\vec{\tilde{\epsilon}} = \mathbf{U}\cdot \vec{\epsilon}$ reads
\begin{align} \label{eq:unitary_2D}
\dunderlines{U} =\left( \begin{array}{ccc} -1/\sqrt{2} & 1/\sqrt{2} & 0 \\ 1/\sqrt{2} & 1/\sqrt{2} & 0 \\ 0 & 0 & 1 \\ \end{array} \right) 
\end{align}
connects the two sets of generators and resulting quantities to each other, as detailed around Eq.~\eqref{eq:linear_transformation}. 
In this way, using our formalism, we readily find the results associated with the types of deformation depicted in Fig.~\ref{fig2} as well.
Evaluating the analog of Eq.~\eqref{eq:small_F} in our two-dimensional setting, we find
\begin{align}
\boldsymbol{\Lambda} 
= \left( \begin{array}{cc} \epsilon_3 & -\epsilon_1 +\epsilon_2 \\ \epsilon_1 +\epsilon_2 & -\epsilon_3 \\ \end{array} \right)
= \left( \begin{array}{cc} \tilde{\epsilon}_3  & \sqrt{2} \tilde{\epsilon}_1 \\ \sqrt{2} \tilde{\epsilon}_2 & -\tilde{\epsilon}_3 \\ \end{array} \right).
\end{align}
%The corresponding deformation gradient tensor is given by
Together with 
\begin{align}
\Feps = \cosh{C_\epsilon}\,\mathbf{I} +\frac{1}{C_\epsilon} \sinh{C_\epsilon}\,\boldsymbol{\Lambda},
\end{align}
$C_\epsilon = \sqrt{{\Lambda}_{11}^2 +{\Lambda}_{12}{\Lambda}_{21}}$, this allows to evaluate Eq.~\eqref{eq:neoHookean_W}. 
%$C_\epsilon = \sqrt{\epsilon_1^2 -\epsilon_2^2 +\epsilon_3^2}$ or $C_\epsilon=\sqrt{2\tilde{\epsilon}_1 \tilde{\epsilon}_2 +{\tilde{\epsilon}_3}^2}$, 
%depending on the generator set used.

Noting that 
%Returning to a predeformation as in Eq.~\eqref{eq_F0}, we note that 
$J = a^2$ and 
setting $\hat{n}_H = (0,1)^T$ and $\hat{n}_s = (0,1)^T$,
%\Feps \cdot \hat{n}_s = \cosh {C_\epsilon} \hat{n}_s + \frac{1}{C_\epsilon}\sinh {C_\epsilon} \vec{l},
%\end{align}
%where $\vec{l} = ({\Lambda}_{12}, {\Lambda}_{22})^T$.
%%Utilizing the series expansions
%%\begin{align}
%%&\cosh{C_\epsilon} = 1+\frac{1}{2}C_\epsilon^2 + O(C_\epsilon^4), \\
%%&\frac{1}{C_\epsilon} \sinh{C_\epsilon} = 1+\frac{1}{6} C_\epsilon^2 + O(C_\epsilon^4),
%%\end{align}
we obtain from Eq.~\eqref{eq:neoHookean_W} up to second order in $\{\epsilon_i\}$ and $\{\tilde{\epsilon}_i\}$, respectively, 
\begin{eqnarray}
W(\{\epsilon_i\})& \approx& \mu (a^2-1-\ln{a^2}) +\frac{\lambda}{2} (\ln{a^2})^2 -\frac{H^2}{2} \nonumber \\
&&{}+2\mu a^2 (\epsilon_2^2 +\epsilon_3^2) +\frac{H^2}{2} (\epsilon_1^2 +\epsilon_2^2 -2\epsilon_1 \epsilon_2)\quad\quad
\end{eqnarray}
and
\begin{eqnarray}
W(\{\tilde{\epsilon}_i\})& \approx &\mu (a^2-1-\ln{a^2}) +\frac{\lambda}{2} (\ln{a^2})^2 -\frac{H^2}{2} \nonumber \\
&&{}+\mu a^2 (\tilde{\epsilon}_1^2 +\tilde{\epsilon}_2^2 +2\tilde{\epsilon}_1 \tilde{\epsilon}_2+ 2\tilde{\epsilon}_3^2) +H^2 \tilde{\epsilon}_1^2. \quad\quad
\end{eqnarray}

On the one hand, comparison with Eqs.~\eqref{eq:expansion} and \eqref{eq:linear_transformation} allows to directly read off $s_i=0=\tilde{s}_i$ ($i=1,2,3$) together with the generalized moduli
%Similarly, the non-zero second-order coefficients are read off as
\begin{align} \label{eq:neoHookean_Cij}
&{C}_{11} = H^2, \quad {C}_{22} = 4\mu a^2 +H^2, \quad {C}_{33} = 4\mu a^2, \nonumber \\
&{C}_{12} ={C}_{21}= -H^2,		
\end{align}
and
\begin{align}
&\tilde{{C}}_{11} = 2\mu a^2 + 2H^2, \quad \tilde{{C}}_{22} = 2\mu a^2, \quad \tilde{{C}}_{33} = 4\mu a^2, \nonumber \\
&\tilde{{C}}_{12} = \tilde{{C}}_{21} = 2\mu a^2.
\end{align}
In this way, the linear transformation rules below Eq.~\eqref{eq:linear_transformation} can directly be verified using Eq.~\eqref{eq:unitary_2D}.
Consequently, this example demonstrates how results for different sets of generators of deformation can readily be obtained from each other.

On the other hand, we may now determine the generalized elastic moduli $C_{ij}$ from our theory using the two-dimensional analogs of Eqs.~(\ref{eq:second_order_infinitesimal}) and (\ref{eq:sigma_structure_constant}). %The structure constants for the associated Lie algebra $\mathfrak{gl}(2,\mathbb{R})$ are listed in \textcolor{red}{Appendix}. 
Specifically, from explicit calculation for the associated Lie algebra $\mathfrak{gl}(2,\mathbb{R})$ with the generators $\boldsymbol{\kappa}_0 \equiv \mathbf{I}$ and $\{\boldsymbol{\kappa}_i\}$ for $i=1,2,3$ given by Eq.~\eqref{eq:kappa_matrices}, we obtain
\begin{align} \label{eq:second_order_infinitesimal_2D}
\frac{1}{2} [\boldsymbol{\sigma}]_{ab} [\{\boldsymbol{\kappa}_i, \boldsymbol{\kappa}_j\}]_{ba}
%&= \frac{1}{2}\left( g^{ij} s_0 +\sum_{k=1}^3 h^{ijk} s_k \right) \nn \\
	= \left( \begin{array}{ccc} 
-s_0 & 0 & 0 \\ 0 & s_0 & 0 \\ 0 & 0 & s_0 \\ \end{array} \right)
_{ij}
%, \quad
.
%\mathbf{C}^2 = \left( \begin{array}{ccc}
%0 & 0 & 0 \\ 0 & 0 & 0 \\ 0 & 0 & 0 \\ \end{array} \right), \quad
\end{align}
For this purpose, we additionally need to calculate $s_0$. As a two-dimensional analog of Eq.~\eqref{eq:symmetric_s0_pressure}, together with the generator $\boldsymbol{\kappa}_0=\mathbf{I}$, we obtain
\begin{align}\label{eq_p2d}
s_0 = {\rm Tr}\,\boldsymbol{\sigma} = 2[\mu (a^2 -1) +\lambda \ln{a^2}] \equiv {}-2p.
\end{align}
%We note that, in our two-dimensional consideration, we find $s_0= {\rm Tr}\,\boldsymbol{\sigma} \equiv -2p$, instead of Eq.~\eqref{eq:symmetric_s0_pressure}, because $\boldsymbol{\kappa}_0 = \mathbf{I}$.
%
Thus, we find in two dimensions
\begin{align} 
{C}_{11} &= [\boldsymbol{\mathcal{A}}]_{1212} + [\boldsymbol{\mathcal{A}}]_{2121}-2[\boldsymbol{\mathcal{A}}]_{1221} - s_0 \nn \\& \equiv B_{1212} +2p, \label{eq:2D_C11} \\
{C}_{22} &= [\boldsymbol{\mathcal{A}}]_{1212} + [\boldsymbol{\mathcal{A}}]_{2121}+2[\boldsymbol{\mathcal{A}}]_{1221} + s_0 \nn \\&\equiv C_{1212} -2p, \label{eq:2D_C22}\\
{C}_{33} &= [\boldsymbol{\mathcal{A}}]_{1111} + [\boldsymbol{\mathcal{A}}]_{2222}-2[\boldsymbol{\mathcal{A}}]_{1122} + s_0, \label{eq:2D_C33} \\
{C}_{12} &= {C}_{21} = -[\boldsymbol{\mathcal{A}}]_{1212}+[\boldsymbol{\mathcal{A}}]_{2121}. \label{eq:2D_C12}
%{C}_{11} &= \mathcal{A}_{1212} + \mathcal{A}_{2121}-2\mathcal{A}_{1221} - s_0 \equiv B_{1212} +2p, \label{eq:2D_C22} \\
%{C}_{22} &= \mathcal{A}_{1212} + \mathcal{A}_{2121}+2\mathcal{A}_{1221} + s_0 \equiv C_{1212} -2p, \label{eq:2D_C11}\\
%{C}_{33} &= \mathcal{A}_{1111} + \mathcal{A}_{2222}-2\mathcal{A}_{1122} + s_0, \label{eq:2D_C33} \\
%{C}_{12} &= {C}_{21} = -\mathcal{A}_{1212}+\mathcal{A}_{2121}. \label{eq:2D_C12}
\end{align}

The newly defined rotation coefficient $B_{1212}$ is linked to rotations generated from $\boldsymbol{\kappa}_1$, see Fig.~\ref{fig1}(a). Likewise, the shear modulus $C_{1212}$ is linked to shear deformations generated from $\boldsymbol{\kappa}_2$, see Fig.~\ref{fig1}(b). They are now directly obtained in an economic way from Eqs.~\eqref{eq:2D_C11} and~\eqref{eq:2D_C22} via Eqs.~\eqref{eq:neoHookean_Cij} and \eqref{eq_p2d}. 
The explicit components of $\boldsymbol{\mathcal{A}}$ are not needed to this end. Yet, they can be calculated from an expansion of $W(\mathbf{F})$ in components of $\Feps$ and using Eq.~\eqref{def:instantaneous} to confirm our expressions.

To summarize our results and illustration, group theory has guided us to an appropriate formulation of nonlinear elasticity under imposed constraints and finite predeformations. Our deformation gradient tensors are constructed consistently from generators, which identifies appropriate expressions. Additional distortions superimposed to finite predeformations, even regarding certain constraints, are in this way described consistently, with consequences even in the infinitesimal limit. Using unitary transformations, the framework is adjusted to the type of deformation at hand. In the limit of infinitesimal superimposed distortions, our theory provides appropriate expressions of generalized elastic moduli and rotation coefficients. 

In addition to our theoretical advance,
it is important to discuss possible applications of the formulation to real systems.
Obviously, our approach should prove (technically) useful in any situation where materials under predeformation or constraints are exposed to external or internal stimuli. 
It will also be illuminating to extend our description to corresponding dynamic scenarios as well as to nonaffine deformations in combination with computational evaluations.
Moreover, in addition to conventional solids, 
various soft and living matters~\cite{Holzapfel2000,Lubensky2002,Storm2005,Mihai2017,Menzel2020} can be modeled as nonlinear elastic materials. 
For example, nematic gels and elastomers \cite{Lubensky2002, Xing2008} can be investigated by our formalism, for which an extension to systems of anisotropic elasticity should be envisaged.

Beyond the technical advance, our approach may open a new avenue to 
investigate nontypical solids. Since the stress vector has been defined as a conjugate to deformation,
our group theoretical approach should be useful to characterize active systems~\cite{Liverpool2009,Mansson2019}, 
in which (active) forces instead of deformations are directly expressed by the system.
While we have discussed our formulation in the context of continuum theory, 
it is straightforward to extend our approach to particle-based models and discretized systems. 
Since an implementation of constraints in this case is not as obvious 
as in continuum models due to fluctuations, our formulation might be relevant for molecular dynamics and Monte-Carlo simulations~\cite{Frenkel2002,Parrinello1982} under constraints and predeformations. 

\begin{acknowledgments}
H.L. was supported by the Deutsche Forschungsgemeinschaft (German Research Foundation, DFG) within the project LO 418/25-1.
A.M.M.\ thanks the Deutsche Forschungsgemeinschaft (German Research Foundation, DFG) for support through the Heisenberg Grant ME 3571/4-1. 
\end{acknowledgments}

\bibliography{elasticity}

\end{document}